\documentstyle{article}

\begin{document}

\title{Linear solutions for cryptographic nonlinear sequence generators}
\date{}
\author{A. F\'{u}ster-Sabater$^{(1)}$ and P. Caballero-Gil$^{(2)}$\\
{\small (1) Instituto de F\'{\i}sica Aplicada, C.S.I.C., Serrano 144, 28006 Madrid, Spain} \\ 
{\small amparo@iec.csic.es}\\
{\small (2) DEIOC, University of La Laguna, 38271 La Laguna, Tenerife, Spain} \\
{\small pcaballe@ull.es }}

\maketitle

\begin{abstract}
This letter shows that linear Cellular Automata based on rules
90/150 generate all the solutions of linear difference equations
with binary constant coefficients. Some of these solutions are
pseudo-random noise sequences with application in cryptography:
the sequences generated by the class of shrinking generators.
Consequently, this contribution shows that shrinking generators do
not provide enough guarantees to be used for encryption purposes.
Furthermore, the linearization is achieved through a simple
algorithm about which a full description is provided.


\begin{center}
"Linearity is the curse of the cryptographer" (J. L. Massey,
Crypto'89)
\end{center}
\end{abstract}

\section{Introduction}
\footnotetext{Final version published in Physics Letters A Vol. 369, Is. 5-6, 1 Oct. 2007, pp. 432-437 \\
DOI:10.1016/j.physleta.2007.04.103  
}
\noindent Secret-key cryptography is commonly divided into block
and stream ciphers. As opposed to block ciphers, stream ciphers
encrypt each data symbol (as small as a bit) into a ciphertext
symbol under a nonlinear dynamical transformation. Stream ciphers
are the fastest among the encryption procedures so they are
implemented in many practical applications e.g. the algorithms A5
in GSM communications \cite{GSM}, the generator RC4 in Wi-Fi
security protocol \cite{Wifi} or the encryption system E0 in
Bluetooth specifications \cite{Bluetooth}.

A stream cipher procedure is based on the generation of a long
keyed Pseudo-random Noise (PN) sequence and its addition to the
original message. In particular, for encryption the sender
realizes the bit-wise XOR operation among the bits of the original
message or plaintext and the pseudo-random noise sequence, giving
rise to the ciphertext. For decryption, the receiver generates the
same pseudo-random noise sequence, realizes the same bit-wise XOR
operation between the received ciphertext and the pseudo-random
noise sequence and recuperates the original message.

Most pseudo-random noise sequence generators are based on either
chaotic encryption procedures (see for instance \cite{Baptista},
\cite{Chee} and \cite{Yu}), or Linear Feedback Shift Registers
(LFSRs) \cite{Golomb}. The output sequences of LFSRs have
application in multiple areas such as spread spectrum
communication, digital ranging, tracking systems, simulation of
random processes, computer sequencing and timing schemes. For
their use in cryptography, such sequences are combined by means of
nonlinear functions. That is the case of combinational generators,
nonlinear filters, clock-controlled generators and irregularly
decimated generators. All of them produce pseudo-random noise
sequences with high linear complexity, long period and good
statistical properties (see \cite{Fuster} and \cite{Caballero}).

Cellular Automata (CA) are discrete structures with dynamical
behaviour extensively studied and applied in modelling systems in
physics, chemistry, biology, computer science and other
disciplines. It has been proved \cite{Cattell} that
one-dimensional linear CA generate exactly the same pseudo-random
noise sequences as those of LFSRs. Regarding more complex
generators, this work shows that certain CA generate exactly the
same pseudo-random noise sequences as those of nonlinear
generators based on LFSRs. Linearity in cipher's behavior, say the
cryptanalysts, is the end of a cipher. It essentially means that
information is leaked from the plaintext to the ciphertext. In
particular, this letter proves how a well known class of
LFSR-based nonlinear generators, the shrinking generators, can be
modelled in terms of linear CA. According to the cryptanalytic
statement, this class of cryptographic generators has been broken.

This contribution proposes the use of difference equations and
cellular automata to predict the dynamic behavior of certain
nonlinear noise sequences. The predictability of such sequences is
carried out through the linearization of their generator.
Furthermore, that linearization process seems to be applicable for
more general noise sequence generators such as those based on
quantum physics and chaotic processes.

\section{The class of shrinking generators}

A shrinking generator is a nonlinear binary sequence generator
composed by two LFSRs (see \cite{Coppersmith}): a control register
notated $R_{1}$ that decimates the sequence produced by the other
register notated $R_{2}$. Let $L_j \in N $; $ (j=1,2)$ be
their corresponding lengths with $(L_1, L_2)=1$ and let $P_j(x)\in
GF(2)[x]\; (j=1,2)$ be their corresponding characteristic
polynomials of degree $L_j$. In practical applications, such
polynomials are primitive in order to generate PN-sequences of
maximum length. Henceforth, $\{a_{i}\}$ and $\{b_{i}\}$ $(i\geq
0)$ $a_{i}, b_{i} \in GF(2)$ denote the binary sequences generated
by $R_{1}$ and $R_{2}$, respectively. The output sequence of the
generator (\textit{the shrunken sequence}) is denoted by
$\{c_{j}\}$ $(j\geq 0)$ with $c_{j} \in GF(2)$. The sequence
produced by $R_{1}$ determines what elements of the sequence
produced by $R_{2}$ are included in the shrunken sequence. The
decimation rule is:

\begin{enumerate}
\item  If $a_{i}=1\Longrightarrow c_{j}=b_{i}$

\item  If $a_{i}=0\Longrightarrow b_{i}$ is discarded.
\end{enumerate}

\noindent A simple example illustrates the behavior of this
structure.

\noindent \textit{Example 1: }Let us consider the following LFSRs:

\begin{enumerate}
\item  $R_{1}$ of length $L_{1}=3$, characteristic polynomial
$P_{1}(x)=1+x^{2}+x^{3}$ and initial state $IS_{1}=(1,0,0)$. The
PN-sequence generated by $R_{1}$ is $\{1,0,0,1,1,1,0\}$ with
period $T_1=2^{L_1}-1=7$.

\item  $R_{2}$ of length $L_{2}=4$, characteristic polynomial
$P_{2}(x)=1+x+x^{4}$ and initial state $IS_{2}=(1,0,0,0)$.
\thinspace The PN-sequence generated by $R_{2}$ is
$\{1,0,0,0,1,0,0,$ $1,1,0,1,0,1,1,1\}$ with period
$T_2=2^{L_2}-1=15$.
\end{enumerate}

The output sequence $\{c_{j}\}$ is given by:

\begin{itemize}
\item  $\{a_{i}\}$ $\rightarrow $ $1\;0\;0\;1\;1\;1\;0\;1\;0\;0\;1\;1\;1\;0%
\;1\;0\;0\;1\;1\;1\;0\;1\;.....$

\item  $\{b_{i}\}$ $\rightarrow $ $\hspace{0.02cm}1\;\underline{0}\;\underline{0}\;0\;1\;0\;%
\underline{0}\;1\;\underline{1}\;\underline{0}\;1\;0\;1\;\underline{1}\;1\;\underline{1}%
        \;\underline{0}\;0\;0\;1\;\underline{0}\;0\;.....$

\item  $\{c_{j}\}$ $\rightarrow $
$1\;0\;1\;0\;1\;1\;0\;1\;1\;0\;0\;1\;0\;.....$
\end{itemize}

According to the decimation rule, the underlined bits
\underline{0} or \underline{1} in $\{b_{i}\}$ are discarded. Thus,
the sequence produced by the shrinking generator is a decimation
of $\{b_{i}\}$ governed by the bits of $\{a_{i}\}$. According to
\cite{Coppersmith}, the period of the shrunken sequence is
$T=(2^{L_{2}}-1)2^{(L_{1}-1)}$ and its linear complexity, notated
$LC$, satisfies the following inequality
\begin{equation}\label{equation:1}
L_{2} \thinspace 2^{(L_{1}-2)}<LC\leq L_{2} \thinspace
2^{(L_{1}-1)}.
\end{equation}
In addition, the shrunken sequence is balanced and has good
distributional statistics. Therefore, this scheme is suitable for
practical implementation of stream cipher cryptosystems and
pattern generators.

\section{Linear multiplicative polynomial CA}
CA are particular forms of finite state machines defined as
uniform arrays of identical cells in an $n$-dimensional space (see
\cite{Kari}). The cells change their states (contents)
synchronously at discrete time instants. The next state of each
cell depends on the current states of the neighbor cells according
to its transition rule. If the transition rules are all linear, so
will be the automaton under consideration. In this letter, we will
deal with a particular kind of binary CA, the so-called linear
multiplicative polynomial cellular automata. They are discrete
dynamical systems characterized by:
\begin{enumerate}
\item Their underlying topology is one-dimensional, that is they
can be represented by a succession of $L$ cells where $L$ is an
integer that denotes the length of the automaton. The state of the
\textit{i-th} cell at instant $n$, notated $x_{i}^{n}$, takes
values in a finite field $x_{i}^{n} \in GF(2)$.

\item They are linear cellular automata as the transition rule for
each cell is a linear mapping $\Phi_i : GF(2)^k\rightarrow GF(2)$
where
\begin{equation}\label{equation:2}
x_{i}^{n+1}=\Phi_i (x_{i-q}^{n},\ldots ,x_{i}^{n},\ldots
,x_{i+q}^{n}) \;\; (i=1, ..., L)
\end{equation}
$k=2q+1$ being the size of the neighborhood.

\item Each one of these cellular automata is uniquely represented
by an $L$ x $L$ transition matrix $M$ over $GF(2)$. The
characteristic polynomial of such matrices is of the form
\begin{equation} \label{equation:3}
 P_M(x)=(P(x))^{p}
\end{equation}
 where $P(x)=x^{r} +
\sum\limits_{j=1}^{r}c_{j} \; x^{r-j}$ denotes a irreducible
(primitive) polynomial of degree $r$ over $GF(2)$ and $p$ an
integer such that $L=p \cdot r$.
\end{enumerate}

This letter is concentrated on one-dimensional binary linear CA
with neighborhood size $k=3$ and particular transition rules
defined as follows:
\begin{center}
Rule 90  \qquad \qquad \qquad \qquad \qquad \qquad Rule 150\\
$x_{i}^{n+1}=x_{i-1}^{n} \oplus x_{i+1}^{n}$ \qquad \qquad \qquad
$x_{i}^{n+1}=x_{i-1}^{n}\oplus x_{i}^{n}\oplus x_{i+1}^{n}$
\end{center}
\noindent where the symbol $\oplus$ represents the XOR logic
operation. Remark that they are linear and very easy transition
rules involving just the addition of either two bits (rule 90) or
three bits (rule 150).

For a cellular automaton of length $L=10$ cells, configuration
rules $(\,90,150,$ $150,150,90,90,150,150,150,90\,)$ and initial
state $(0,0,0,1,1,1,0,1,1,0)$, Table \ref{table:1} illustrates the
behavior of this structure: the formation of its output sequences
(binary sequences read vertically) and the succession of states
(binary configurations of 10 bits read horizontally). In addition,
cells with permanent null contents are supposed to be adjacent to
the array extreme cells.

The characteristic polynomial $P(x)$ of an arbitrary binary
sequence $\{a_n\}$ specifies its linear recurrence relationship.
This means that the \textit{n-th} element $a_n$ can be written as
a linear combination of the previous elements:
\begin{equation}\label{equation:4}
a_{n}\oplus \sum\limits_{i=1}^{r}c_{i} \; a_{n-i}=0,\qquad n\geq
r.
\end{equation}
The linear recursion is expressed as a linear difference equation:
\begin{equation}\label{equation:5}
(E^{r}\oplus \sum\limits_{i=1}^{r}c_{i}\;E^{r-i})\;a_{n}=0,\qquad
n\geq 0
\end{equation}
where $E$ is the shifting operator that operates on $a_n$, i.e.
$Ea_n=a_{n+1}$. If the characteristic polynomial $P(x)$ is
primitive and $\alpha$ one of its roots, then
\begin{equation}\label{equation:6}
\alpha, \;\alpha^2, \;\alpha^{2^2}, \ldots, \;\alpha^{2^{(r-1)}}
\end{equation}
are the $r$ different roots of such a polynomial as well as
primitive elements in $GF(2^r)$ (see \cite{Lidl}).

Now, if the characteristic polynomial of of an arbitrary binary
sequence $\{a_n\}$ is of the form $P_M(x)=(P(x))^{p}$ as defined
in (\ref{equation:3}), then its roots will be the same as those of
$P(x)$ but with multiplicity $p$. The corresponding difference
equation will be:
\begin{equation}\label{equation:7}
(E^{r}\oplus \sum\limits_{i=1}^{r}c_{i}\;E^{r-i})^p \;
a_{n}=0,\qquad n\geq 0
\end{equation}
and its solutions are of the form
$a_{n}=\sum\limits_{j=0}^{r-1}\;\sum\limits_{m=0}^{p-1}({n \choose m}\thinspace
A_m^{2^j}) \; \alpha^{2^jn}$, where $A_m$ is an arbitrary element
in $GF(2^L)$. Different choices of $A_m$ will give rise to
different sequences $\{a_n\}$. Consequently, all the binary
sequences $\{a_n\}$ of characteristic polynomial
$P_M(x)=(P(x))^{p}$ can be generated by linear multiplicative
polynomial CA as well as all of them are solutions of the linear
difference equation described in (\ref{equation:7}). Our analysis
focuses on all the possible solutions of this equation.

\section{Realization of linear multiplicative polynomial
CA}

In the previous section, algebraic properties of the sequences
obtained from multiplicative polynomial CA have been considered.
Now the particular form of these automata is analyzed.

A natural way of representation for this type of 90/150 linear CA
is a binary \textit{L}-tuple $ \Delta=(d_1, d_2, ..., d_L)$ where
$d_i=0$ if the \textit{i-th} cell verifies rule 90 while $d_i=1$
if the \textit{i-th} cell verifies rule 150. The Cattell and Muzio
synthesis algorithm \cite{Cattell} presents a method of computing
two 90/150 CA corresponding to a given polynomial. Such an
algorithm takes as input an irreducible polynomial $Q(x)$ and
computes two reversal \textit{L}-tuples corresponding to two
different linear CA whose output sequences have $Q(x)$ as
characteristic polynomial. The total number of operations required
for this algorithm is linear in the degree of the polynomial and
is listed in \cite{Cattell}(Table II, page 334). The method is
efficient for all practical applications (e.g. in 1996 finding a
pair of length $300$ CA took 16 CPU seconds on a SPARC 10
workstation). For cryptographic applications, the degree of the
primitive polynomial $P(x)$ is $ L_{2}\approx 64$, so that the
consuming time is negligible. Finally, a list of one-dimensional
linear CA of degree through 500 can be found in \cite{Cattell1}.

Since the characteristic polynomials we are dealing with are of
the form $P_M(x)= (P(x))^{p}$, it seems quite natural to construct
a multiplicative polynomial cellular automaton by concatenating
$p$ times the automaton whose characteristic polynomial is $P(x)$.
The procedure of concatenation is based on the following result.

{\bf Lemma 1.}
Let $\Delta=(d_1, d_2, ...,d_L)$ be the representation of an
one-dimensional binary linear cellular automaton with $L$ cells
and characteristic polynomial $P_L(x)=(x+d_1)(x+d_2)...(x+d_L)$.
The cellular automaton whose characteristic polynomial is
$P_{2L}(x)=(P_L(x))^2$ is represented by:
\begin{equation}\label{equation:8}
\Delta=(d_1, d_2, ..., \overline{d_L}, \overline{d_L}, ..., d_2,
d_1)
\end{equation}
where the overline symbol represents bit complementation.

\textit{Proof.}$\;$ The result follows from the fact that:
\[
P_{\overline{L}}(x) =P_L(x)+P_{L-1}(x)
\]
where $P_{\overline{L}}(x)$ is the polynomial corresponding to
$\Delta=(d_1, d_2, ..., \overline{d_L})$. In the same way
\[
P_{L+1}(x) =(x+d_L)P_{\overline{L}}(x)+P_{L}(x)
\]
\[
P_{L+2}(x) =(x+d_{L-1})P_{L+1}(x)+P_{\overline{L}}(x)
\]
\[
\vdots \quad \quad \quad \quad \vdots
\]
\[
P_{2L}(x) =(x+d_1)P_{2L-1}(x)+P_{2L-2}(x).
\]
Thus, by successive substitutions of the previous polynomial into
the next one we get:
\begin{equation}\label{equation:9}
P_{2L}(x) =(x+d_1)P_{2L-1}(x)+P_{2L-2}(x)=(P_L(x))^2.
\end{equation}\hfill $\Box$

The result can be iterated for successive exponents. In this way,
the concatenation of an automaton and its mirror image allows us
to realize linear multiplicative polynomial CA. The
complementation is due to the fact that rule $90$ ($150$) at the
end of the array is equivalent to two consecutive rules $150$
($90$) with identical sequences.

\section{Shrunken sequences as solutions of linear equations: a simple linearization procedure}
Now the result that relates the shrunken sequences from shrinking
generators with the sequences obtained from linear multiplicative
polynomial cellular automata is introduced.

{\bf Theorem 1.}
The characteristic polynomial of the output sequence of a
shrinking generator with parameters $L_j\in N$ and
$P_j(x)\in GF(2)[x]\; (j=1,2)$ defined as in section (2) is of the
form $P_M(x)= (P(x))^{p}$, where $P(x)\in GF(2)[x]$ is a
$L_{2}$-degree polynomial and $p$ is an integer satisfying the
inequality $2^{(L_{1}-2)}<p\leq 2^{(L_{1}-1)}$.

Proof. The shrunken sequence can be written
as a sequence made out of an unique \textit{PN}-sequence starting
at different points and repeated $2^{(L_{1}-1)}$ times. Such a
sequence is obtained from $\{b_{i}\}$ taking elements separated a
distance $2^{L_1}-1$, that is the period of the sequence
$\{a_{i}\}$. As $(2^{L_2}-1, 2^{L_1}-1)=1$ due to the primality of
$L_2$ and $L_1$, the result of the decimation of $\{b_{i}\}$ is a
\textit{PN}-sequence whose characteristic polynomial $P(x)$ of
degree $L_2$ is the characteristic polynomial of the
\textit{cyclotomic coset} $2^{L_{1}}-1$, that is $P(x)
=(x+\alpha^{N})(x+\alpha^{2N})\ldots (x+\alpha^{2^{L_{1}-1}N})$
being $N$ an integer given by $N =2^{0}+2^{1}+\ldots
+2^{L_{1}-1}$. Moreover, the number of times that this
\textit{PN}-sequence is repeated coincides with the number of
$1's$ in $\{a_{i}\}$ since each $1$ of $\{a_{i}\}$ provides the
shrunken sequence with $2^{L_2}-1$ elements of $\{b_{i}\}$.
Consequently, the characteristic polynomial of the shrunken
sequence will be $P(x)^{p}$ with $p\leq 2^{(L_{1}-1)}$. The lower
limit follows immediately from equation (\ref{equation:3}) and the
definition of linear complexity of a sequence as the shortest
linear recurrence relationship.
\hfill $\Box$

According to its characteristic polynomial, the output sequence of
a shrinking generator is a particular solution of a linear
difference equation as well as it can be generated by linear
multiplicative polynomial CA. Now, the construction of such linear
models from the shrinking generator parameters is carried out by
the following algorithm:

\noindent \textit{Linearization algorithm}

 \textbf{Input:} A shrinking
generator characterized by two LFSRs, $R_{1}$ and $R_{2}$, with
their corresponding lengths, $L_{1}$ and $L_{2}$, and the
characteristic polynomial $P_{2}(x)$ of the register $R_{2}$.

\begin{description}
\item [Step 1] From $L_{1}$ and $P_{2}(x)$, compute the polynomial
$P(x)$ as
\[P(x) =(x+\alpha^{N})(x+\alpha^{2N})\ldots
(x+\alpha^{2^{L_{2}-1}N})\]
with $N =2^{0}+2^{1}+\ldots +2^{L_{1}-1}$.

\item [Step 2] From $P(x)$, apply the Cattell and Muzio synthesis
algorithm to determine two linear 90/150 CA, notated $s_i$, whose
characteristic polynomial is $P(x)$.

\item [Step 3] For each $s_i$ separately, proceed:
\begin{description}
  \item [3.1] Complement its least significant bit. The resulting binary string is notated $S_i$.

  \item [3.2] Compute the mirror image of $S_i$, notated $S_i^{*}$, and
  concatenate both strings
  \[
  S'_{i} = S_i*S_i^{*}\;.
  \]
  \item [3.3] Apply steps $3.1$ and $3.2$ to each $S'_{i}$ recursively $L_{1}-1$ times.
\end{description}

\end{description}

\textbf{Output: } Two binary strings of length $L=L_2 \cdot
2^{L_{1}-1}$ codifying two CA corresponding to the given shrinking
generator.

$ Remark 1.$
In this algorithm the characteristic polynomial of the register
$R_{1}$ is not needed. Thus, all the shrinking generators with the
same $R_2$ but different registers $R_1$ (all of them with the
same length $L_1$) can be modelled by the same pair of
one-dimensional linear CA.

$Remark 2.$
It can be noticed that the computation of both CA is proportional
to $L_{1}$ concatenations. Consequently, the algorithm can be
applied to shrinking generators in a range of practical
application.

$Remark 3.$
In contrast to the nonlinearity of the shrinking generator, the
CA-based models that generate the shrunken sequence are linear.

In order to illustrate the previous steps a numerical example is
presented.

\noindent \textit{Example 2:}

\textbf{Input:} A shrinking generator characterized by two LFSRs:
$R_{1}$ of length $L_{1}=3$, $R_{2}$ of length $L_{2}=5$ and
characteristic polynomial $P_{2}(x)=1+x+x^{2}+x^{4}+x^{5}$.

\begin{description}
\item [Step 1] $P(x)$ is the characteristic polynomial of the
cyclotomic \textit{coset} $N=7$. Thus,
\[
P(x) =1+x^{2}+x^{5}\;.
\]

\item [Step 2] From $P(x)$ and applying the Cattell and Muzio
synthesis algorithm, two reversal linear CA whose characteristic
polynomial is $P(x)$ can be determined. Such CA are written in
binary format as:

\begin{center}
$
\begin{array}{lllll}
0 & 1 & 1 & 1 & 1 \\
1 & 1 & 1 & 1 & 0
\end{array}
$
\end{center}

\item  \textit{Step 3:} Computation of the required pair of CA by
successive concatenations.

For the first automaton:
\begin{center}
$
\begin{array}{llllllllllllllllllll}
0 & 1 & 1 & 1 & 1 &   &   &   &   &   &   &   &   &   &   &   &   &   &   & \\
0 & 1 & 1 & 1 & 0 & 0 & 1 & 1 & 1 & 0 &   &   &   &   &   &   &   &   &   & \\
0 & 1 & 1 & 1 & 0 & 0 & 1 & 1 & 1 & 1 & 1 & 1 & 1 & 1 & 0 & 0 & 1 & 1 & 1 & 0\;\\ 
\end{array} \newline (final \ automaton)
$
\end{center}

For the second automaton:
\begin{center}
$
\begin{array}{llllllllllllllllllll}
1 & 1 & 1 & 1 & 0 \\
1 & 1 & 1 & 1 & 1 & 1 & 1 & 1 & 1 & 1\\
1 & 1 & 1 & 1 & 1 & 1 & 1 & 1 & 1 & 0 & 0 & 1 & 1 & 1 & 1 & 1 & 1
& 1 & 1 & 1\;\\ 
\end{array}\newline(final \ automaton)
$
\end{center}

For each automaton, the procedure of concatenation has been
carried out $L_{1}-1$ times.
\end{description}

\textbf{Output: } Two binary strings of length $L = L_2 \cdot
2^{(L_{1}-1)}=20$ codifying the required pair of CA.

In this way, we have obtained a pair of linear CA:

(90,150,150,150,90,90,150,150,150,150,150,150,150,150,90,90,150,
150,150,90)\\(150,150,150,150,150,150,150,150,150,90,90,150,150,150,150,150,150,150,150,150)

\noindent both of them able to generate the shrunken sequence
corresponding to the given shrinking generator. Consequently, the
shrinking generator can be expressed in terms of a lineal model
based on CA.

\section{Conclusions}
The pseudo-random noise sequence produced by a shrinking generator
is a particular solution of a linear difference equation and can
be generated by linear multiplicative polynomial cellular
automata. In this way, cryptographic generators conceived and
designed as nonlinear generators can be linearized in terms of
cellular automata, which implies that such cryptographic
generators have been broken. The used linearization algorithm is
simple and might be applied to more general sequence generators
such as those based on quantum physics and chaotic processes. \\

\noindent\textbf{Acknowledgements}

 \noindent This work has been supported by
Ministerio de Educaci\'{o}n y Ciencia (Spain), Projects
SEG2004-02418 and SEG2004-04352-C04-03.

\bigskip
\begin{table}[ht]
\caption{An one-dimensional linear cellular automaton of $10$
cells with rules 90/150 starting at a given initial state. The
period of these sequences is $T=62$} \label{table:1}
\renewcommand\arraystretch{1.5}
\noindent\[
\begin{tabular}{cccccccccc}
\hline\noalign{\smallskip}
$\;\;90\;$ & $\;\;150$ & $\;\;150$ & $\;\;150$ & $\;\;90\;$ & $\;\;90\;$ & $\;\;150$ & $\;\;150$ & $\;\;150$ & $\;\;90\;$ \\
\noalign{\smallskip} \hline \noalign{\smallskip} \hline
$\;0$ & $\;0$ & 0 & 1 & 1 & 1 & 0 & 1 & 1 & 0 \\
$\;0$ & $\;0$ & 1 & 0 & 0 & 1 & 0 & 0 & 0 & 1 \\
$\;0$ & $\;1$ & 1 & 1 & 1 & 0 & 1 & 0 & 1 & 0 \\
$\;1$ & $\;0$ & 1 & 1 & 1 & 0 & 1 & 0 & 1 & 1 \\
$\;0$ & $\;0$ & 0 & 1 & 1 & 0 & 1 & 0 & 0 & 1 \\
$\;0$ & $\;0$ & 1 & 0 & 1 & 0 & 1 & 1 & 1 & 0 \\
$\;\vdots$ & $\;\vdots$ & \vdots & \vdots & \vdots & \vdots & \vdots & \vdots & \vdots & \vdots \\
\hline
\end{tabular}
\]
\end{table}

\end{document}